\documentclass[11pt]{article}
\pdfoutput=1
\usepackage{textcomp}
\usepackage{tabularx} 
\usepackage{amsmath}  
\usepackage{graphicx} 
\usepackage{subcaption}
\usepackage[top=2.5cm, bottom=2.5cm, left=2.5cm, right=2.5cm]{geometry}
\usepackage[final]{hyperref} 
\usepackage{titling}
\usepackage{gensymb}
\usepackage[export]{adjustbox}

\hypersetup{
	colorlinks=true,       
	linkcolor=black,        
	citecolor=blue,        
	filecolor=magenta,     
	urlcolor=blue         
}
\usepackage{listings}
\usepackage{color}

\definecolor{dkgreen}{rgb}{0,0.6,0}
\definecolor{gray}{rgb}{0.5,0.5,0.5}
\definecolor{mauve}{rgb}{0.58,0,0.82}

\usepackage[style=ieee, sorting=none, backend=biber, doi=false, isbn=false, maxnames=4, url=false, citestyle=numeric-comp, dashed=false]{biblatex}
\newcommand*{\bibtitle}{} 

\begin{document}

\setlength{\droptitle}{-7em}  
\title{\Large \textbf{DUNE-PRISM: Reducing neutrino interaction model dependence with a movable neutrino detector} \\[0.4em]
\large Contribution to the 25th International Workshop on Neutrinos from Accelerators\\ \vspace{-1.5em}}
\author{\textbf{Ciaran Hasnip}\thanks{ciaran.mark.hasnip@cern.ch}  -- for the DUNE Collaboration\\ \vspace{-1.5em}
\textit{CERN}}
\date{}
\maketitle

\vspace{-4.5em}
\begin{abstract}
\noindent The Deep Underground Neutrino Experiment (DUNE) is a next-generation long-baseline neutrino oscillation experiment designed to make precision measurements in the world's most powerful neutrino beam. Neutrinos are measured at two detector facilities: a near detector (ND) located at Fermilab close to where the beam is produced and a far detector (FD) at SURF. The Precision Reaction Independent Spectrum Measurement (PRISM) system allows for the measurement of different neutrino energy spectra by moving the near detector away from the central axis of the neutrino beam. These off-axis neutrino energy spectra provide a new degree of freedom that can be used to develop a deeper understanding of the relationship between the observable energy deposits in the detector and the energy of the interacting neutrino. This can benefit DUNE by significantly reducing the impact of systematic uncertainties in the neutrino interaction model. One possible use of the PRISM system is to perform a novel neutrino oscillation analysis that linearly combines off-axis neutrino energy spectra at the near detector to produce data-driven predictions of the far detector energy spectrum.
\end{abstract}

\vspace{-1.5em}
\section{\label{sec:intro}Challenges in Modelling Neutrino Interactions}
\vspace{-0.75em}

The large size of the underground detectors at SURF and power of its neutrino beam means that DUNE will be a systematically-limited experiment and uncertainties in the neutrino interaction model present one of the most problematic sources of systematic uncertainties~\cite{DUNE_FDTDR2}.

Oscillation experiments measure the neutrino mixing parameters through measurements of the oscillation probability, which is a function of the neutrino energy. However, experiments observe event rates as a function of reconstructed energy calculated from observable particles in the detector. The neutrino interaction model is therefore necessary to provide a mapping between the kinematics of observable particles produced by a neutrino interaction and the neutrino energy. Any mistake in this mapping could feed-down to an inaccurate measurement of the neutrino mixing parameters.
\vspace{-0.75em}
\section{\label{sec:larsoft}DUNE Near Detector}
\begin{figure}[]
	\centering
    \begin{subfigure}{0.49\textwidth}
	   \includegraphics[width=0.99\columnwidth]{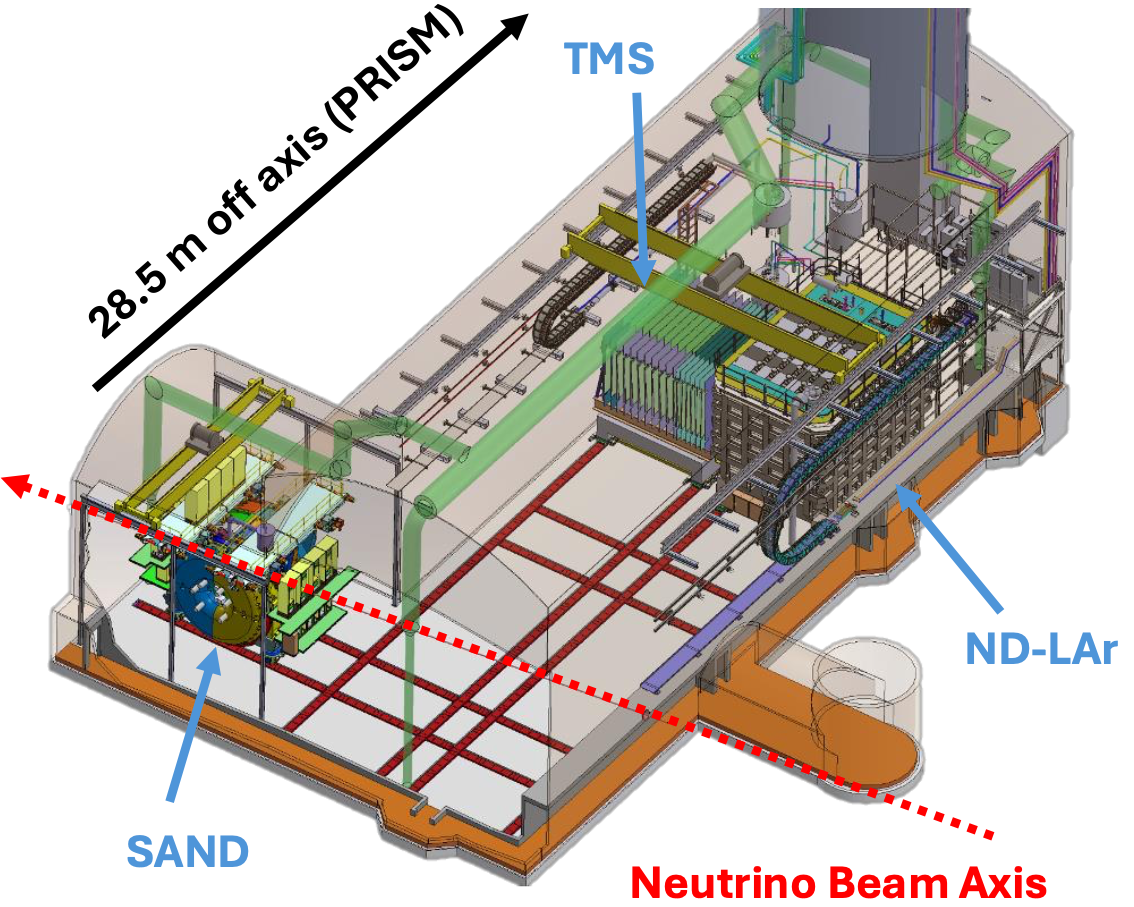}
	   \caption{}
	   \label{fig:DUNEND}
    \end{subfigure}
    \hfill
    \begin{subfigure}{0.45\textwidth}
	   \includegraphics[width=0.90\columnwidth]{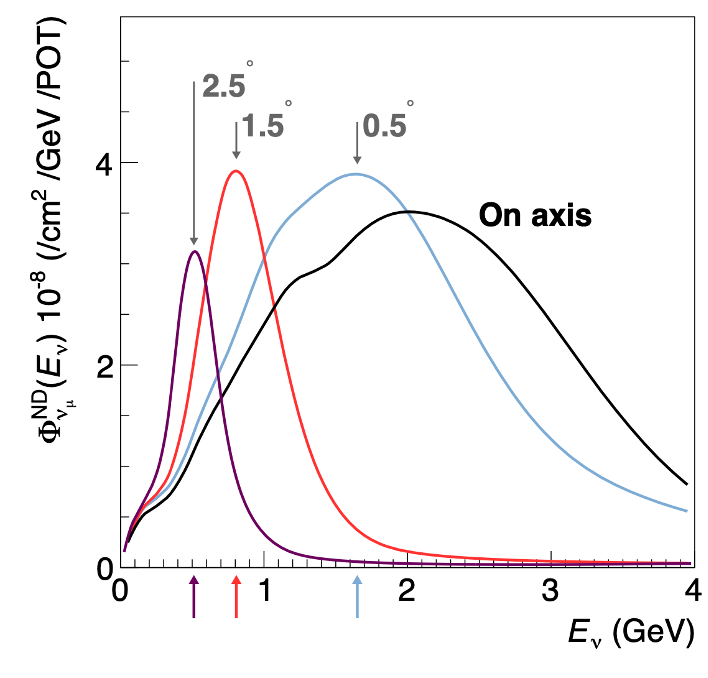}
	   \caption{}
	   \label{fig:flux}
    \end{subfigure}
    \caption{Diagram of the ND facility (left). The ND-LAr and TMS detectors can move 28.5~m off the axis of the neutrino beam. The right plot (Fig.~\ref{fig:flux}~\cite{DUNENDCDR}) shows the flux measured by ND-LAr at different off-axis positions. An off-axis angle of $2.5\degree$ corresponds to a 25~m displacement.}
    \label{fig:dunendflux}
\end{figure}

\vspace{-0.75em}
The DUNE Near Detector (ND) is a multi-component detector complex located 574~m downstream from the LBNF target. A schematic diagram of the ND facility is shown in Fig.~\ref{fig:DUNEND} in which the three detectors are labelled. The purpose of the ND is to measure a large number of neutrinos before oscillation in order to constrain systematic uncertainties.

The most upstream component of the ND is ND-LAr; a modular liquid argon time projection chamber (LArTPC) design. Downstream of ND-LAr is the Temporary Muon Spectrometer (TMS), which is an iron and scintillator bar tracker that measures the energy of muons exiting ND-LAr from $\nu_{\mu}$/$\bar{\nu}_{\mu}$ interactions. The third component is the System for on-Axis Near Detection (SAND), which acts as a neutrino flux monitor and can contribute to cross-section measurements~\cite{DUNENDCDR}.

The final feature of the ND is the ability to continuously move ND-LAr and TMS horizontally off the central axis of the neutrino beam by up to 28.5~m. This capability is called the Precision Reaction Independent Spectrum Measurement (PRISM). The muon neutrino flux narrows and moves to lower energies as the detector moves further off axis, as can be seen in Fig.~\ref{fig:flux}.
\vspace{-0.75em}
\section{\label{sec:prismlc}PRISM Linear Combination}
\vspace{-0.75em}
There are two approaches to using off-axis fluxes. The off-axis measurements could be included in a comparison between ND data and Monte Carlo (MC) to better-constrain systematic uncertainties. This will incorporate an improved understanding of the relationship between the observable energy deposits and the neutrino energy into the measurement of the neutrino mixing parameters. However, no complete model of neutrino-nuclei interactions exists and model tuning and empirical corrections are likely to still be needed. PRISM presents an opportunity to take a different approach to measuring neutrino oscillations, where the far detector (FD) event rate of oscillated neutrinos is predicted directly from the ND data~\cite{nuPRISM_2014, Hasnip2023y}.

\begin{figure}[t]
  \begin{subfigure}{0.49\textwidth}
  \includegraphics[width=0.99\linewidth]{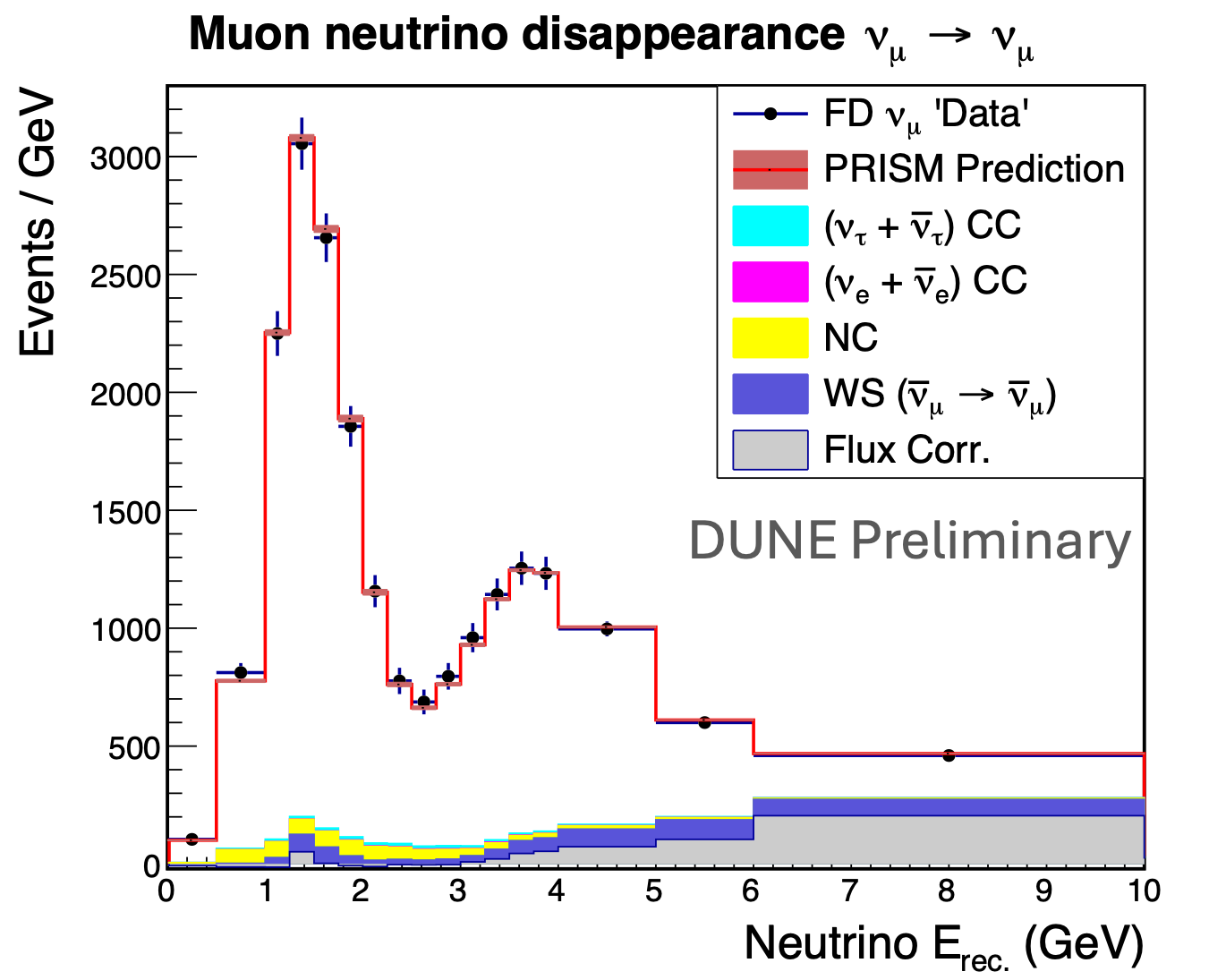}
  \end{subfigure}
  \hfill
  \begin{subfigure}{0.49\textwidth}
    \includegraphics[width=0.99\linewidth]{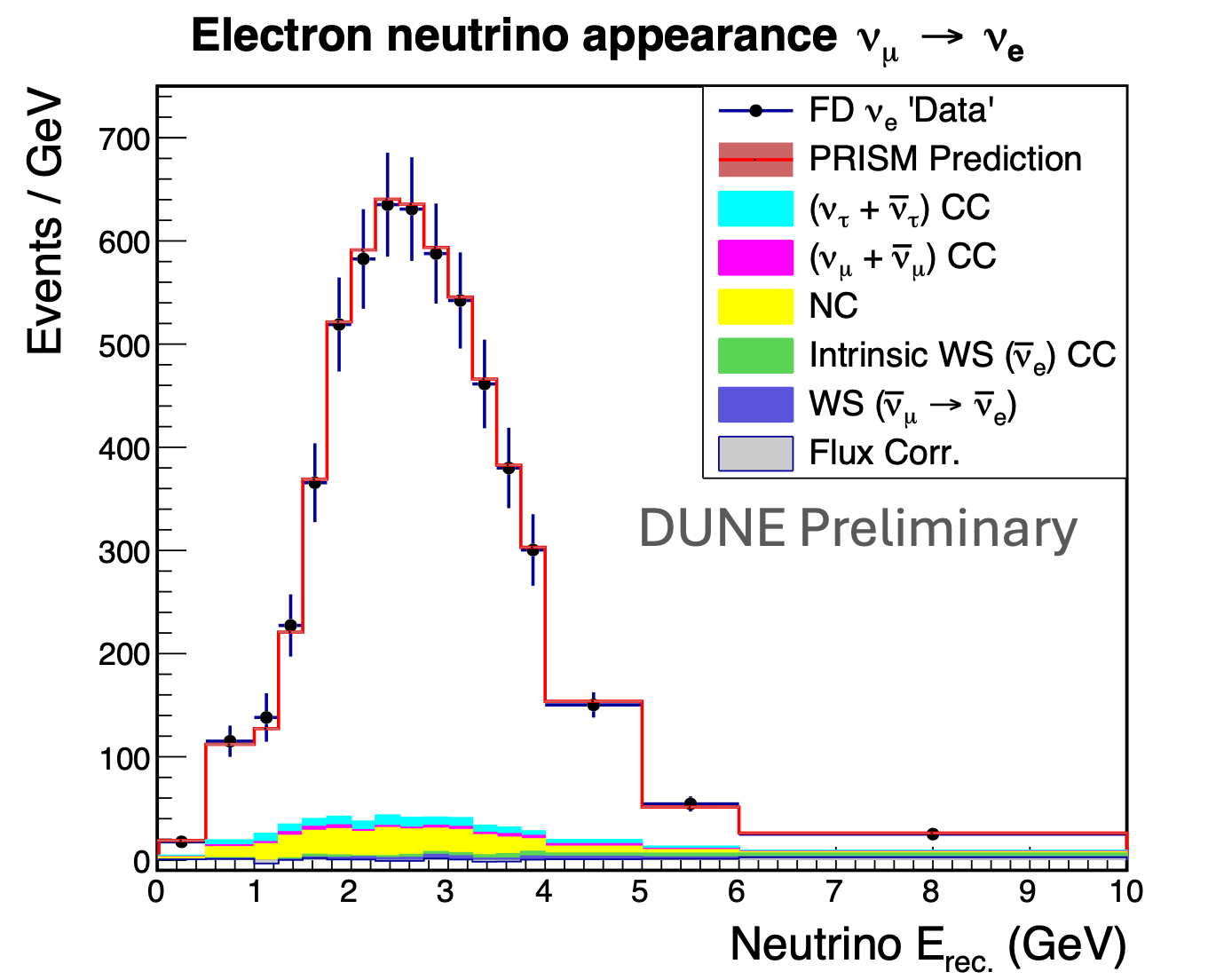}
  \end{subfigure}
  
  \begin{subfigure}{0.49\textwidth}
  \includegraphics[width=0.99\linewidth]{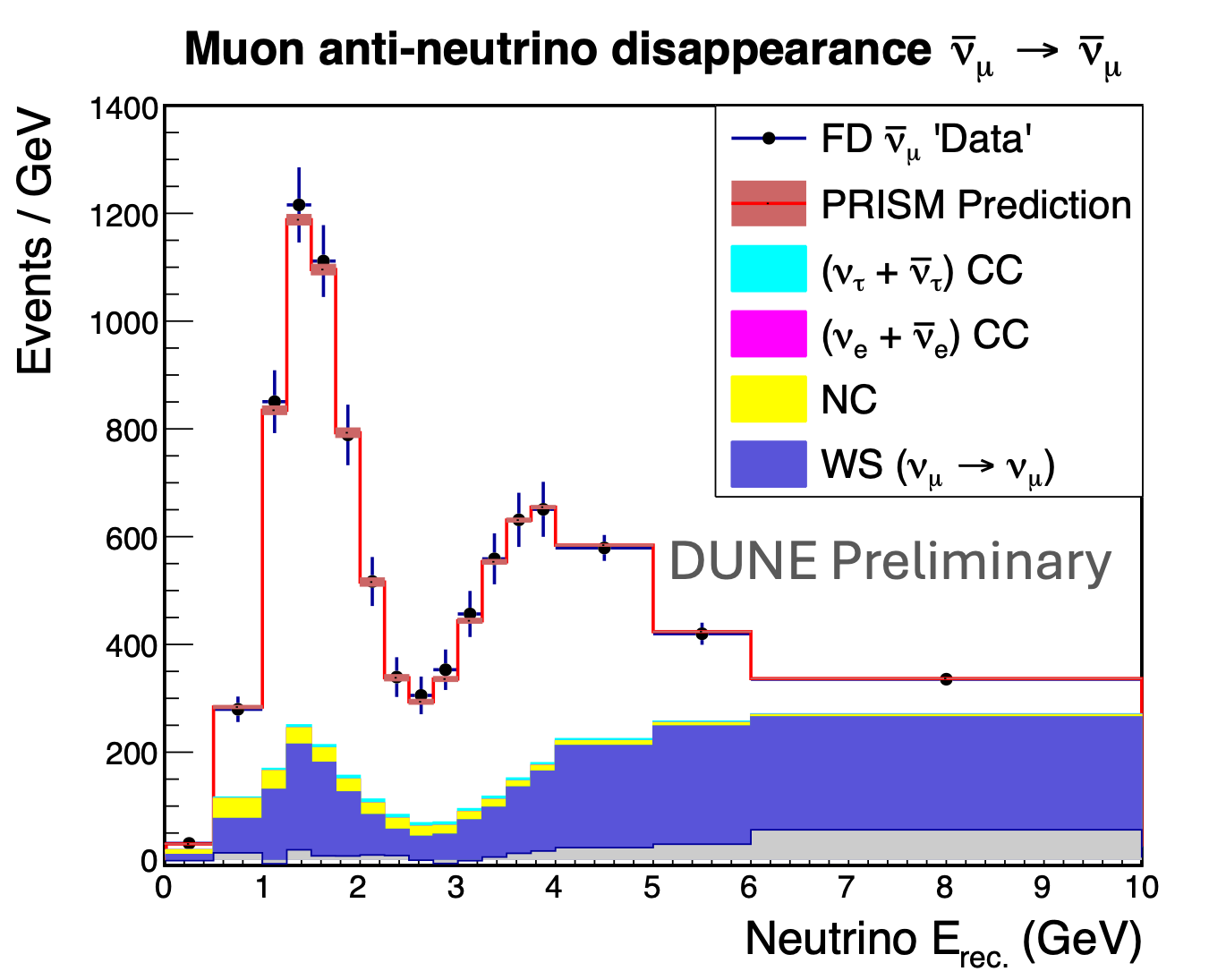}
  \end{subfigure}
  \hfill
  \begin{subfigure}{0.49\textwidth}
    \includegraphics[width=0.99\linewidth]{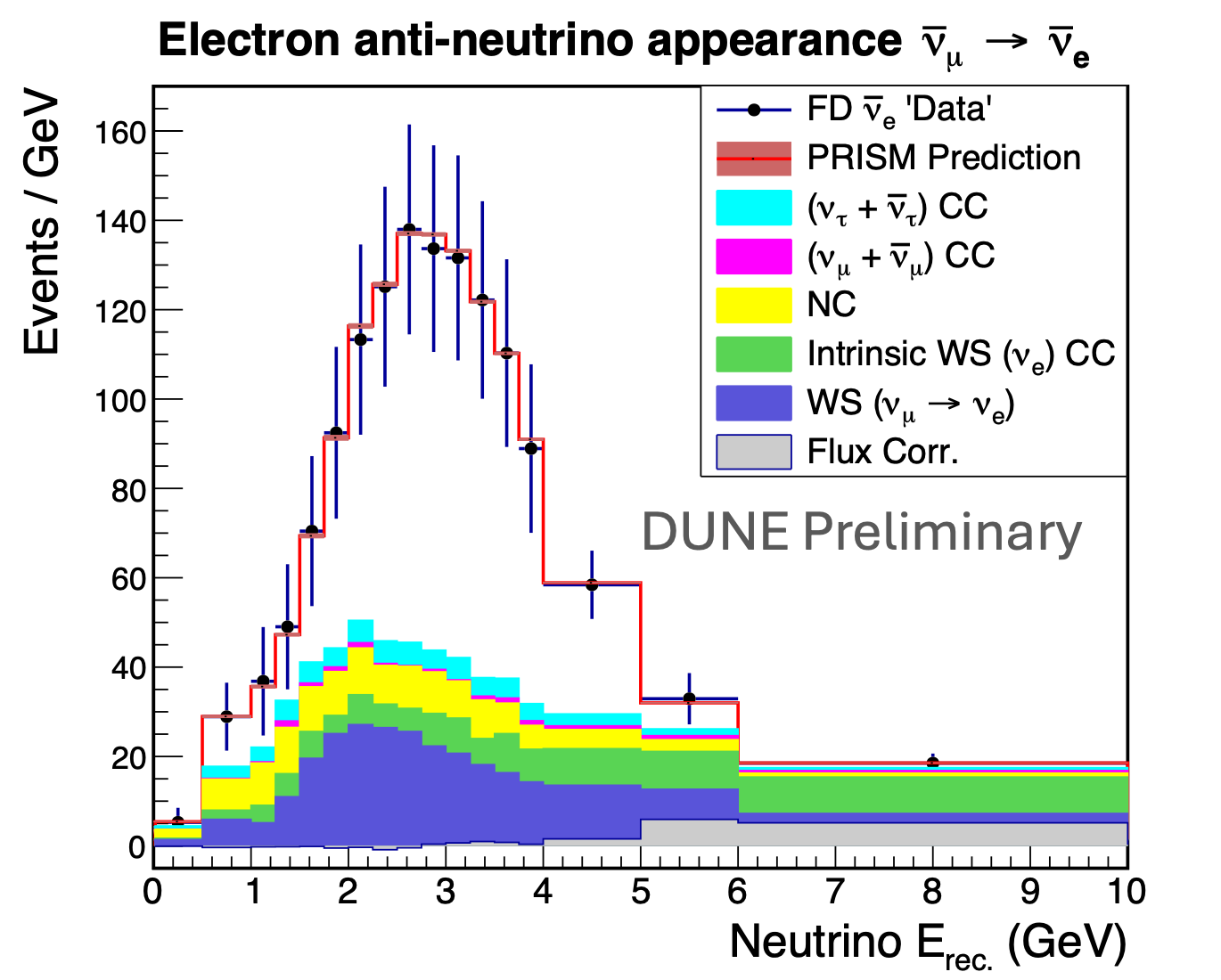}
  \end{subfigure}
  
  \caption{PRISM linear combination predictions for the four oscillation channels. The total stacked histogram is the PRISM prediction, where the solid colour components are MC-derived corrections, mostly for the FD backgrounds. The red component is the linear combination of ND data.}
  \label{fig:prismpreds}
\end{figure}

A prediction of the DUNE FD event rate is made by linearly combining the off-axis fluxes measured by the moveable ND. By building the FD event rate prediction directly from data the correct neutrino interaction physics is naturally incorporated into the DUNE neutrino oscillation measurement. The predicted FD event rate in reconstructed energy bin $j$ can be calculated as
\begin{equation}
  F_{j}^{LC} = \sum_{i}^{n_{pos}} N_{ij}^{data}c_{i},
  \label{eq:lc}
\end{equation}
where $N_{ij}^{data}$ is the ND event rate in reconstructed energy bin $j$ and off-axis position $i$ and $c_{i}$ is a coefficient associated with a particular off-axis position $i$. There are $n_{pos}$ off-axis positions in total. The ND data are $\nu_{\mu}$ ($\bar{\nu}_{\mu}$) event rates measured in neutrino (anti-neutrino) beam mode at each off-axis position. The ND event rates in Eq.~\ref{eq:lc} have been corrected for backgrounds and efficiency and resolution differences between the ND and FD. The coefficients are calculated by mapping ND off-axis fluxes to a target FD oscillated flux. This calculation only uses the flux simulation and is entirely neutrino interaction model independent. Examples of some PRISM linear combination predictions for the four channels measured by DUNE are shown in Fig.~\ref{fig:prismpreds}. Each figure shows the total PRISM prediction and the simulated FD event rate that is being predicted.

\vspace{-0.75em}
\section{\label{sec:prismosc}PRISM Oscillation Analysis}
\vspace{-0.75em}
PRISM predictions (Fig.~\ref{fig:prismpreds}) can be produced for any neutrino mixing parameters by altering the target FD oscillated flux when calculating the linear combination coefficients $c$. A PRISM oscillation measurement is performed by defining equidistant points in an oscillation parameter space and evaluating the goodness of fit between the PRISM prediction and FD event rate at each point. This results in a surface of likelihood values from which DUNE sensitivity contours are drawn.

\begin{figure}[]
	\centering
    \begin{subfigure}{0.43\textwidth}
	   \includegraphics[width=0.99\columnwidth]{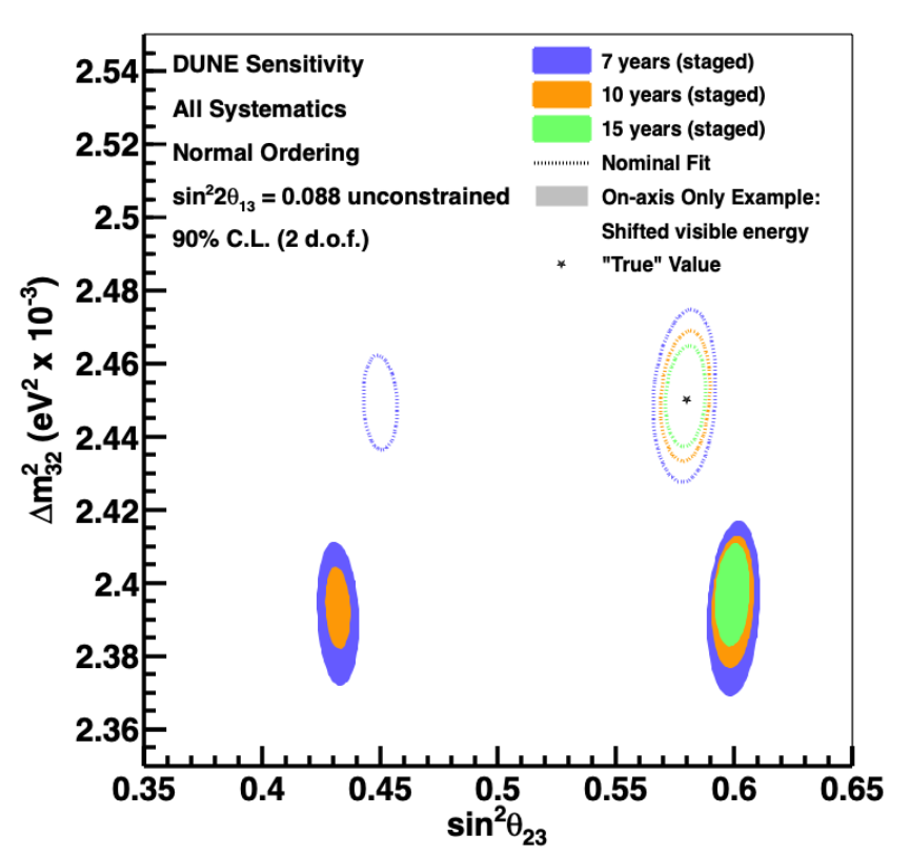}
	   \caption{}
	   \label{fig:cdrcont}
    \end{subfigure}
    \hfill
    \begin{subfigure}{0.49\textwidth}
	   \includegraphics[width=0.99\columnwidth]{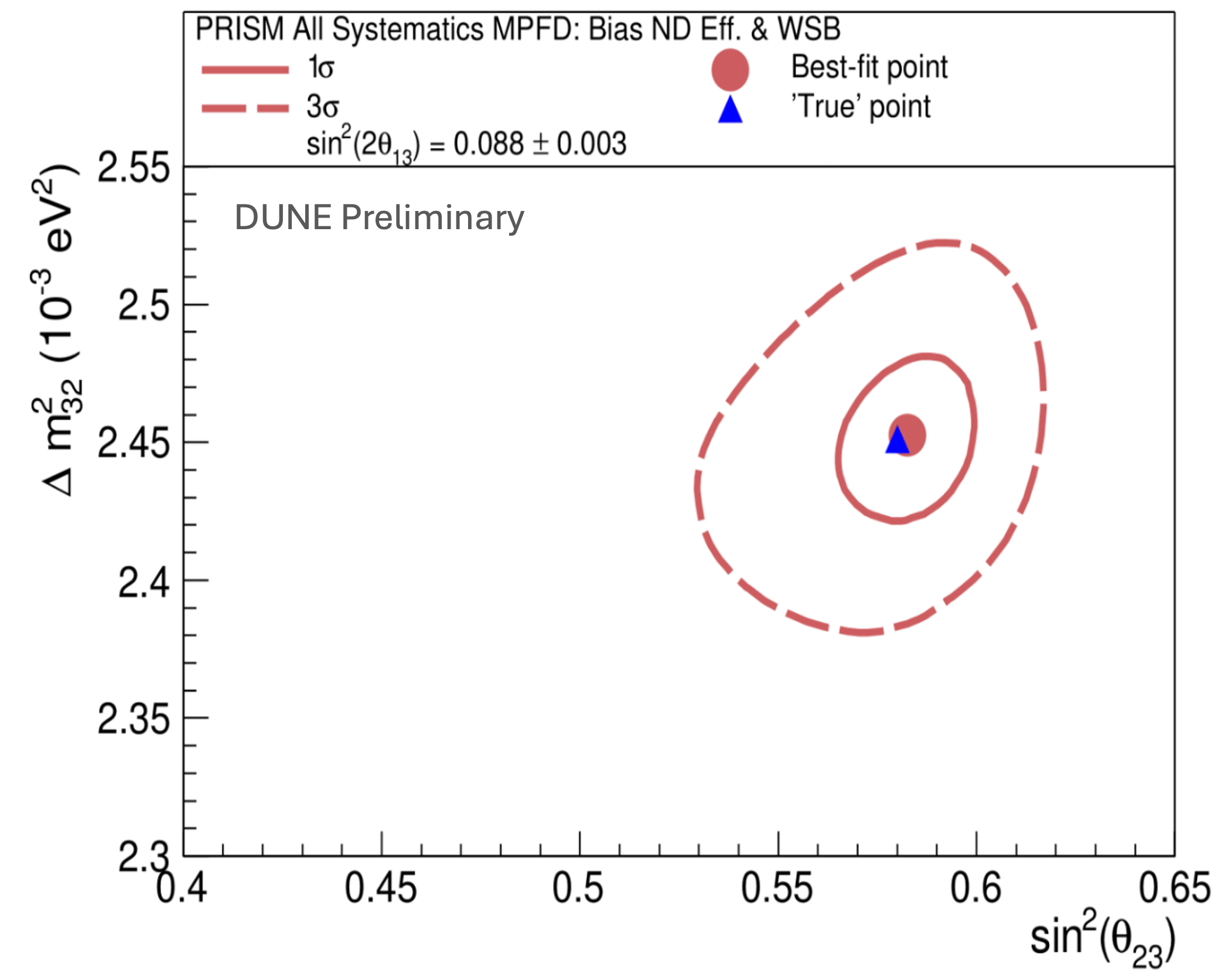}
	   \caption{}
	   \label{fig:prismcontour}
    \end{subfigure}
    \caption{DUNE measurement of $\theta_{23}$ and $\Delta m^{2}_{32}$ using the model dependent (Fig.~\ref{fig:cdrcont}~\cite{DUNENDCDR}) and PRISM linear combination (Fig.~\ref{fig:prismcontour}~\cite{Hasnip2023y}) methods. The bias is no longer present in PRISM method case.}
    \label{fig:contours}
\end{figure}

Fig.~\ref{fig:contours} presents a case study of a DUNE measurement in which the neutrino interaction model present in the MC does not accurately describe the relationship between the observable energy and neutrino energy. This is despite there appearing to be good agreement between data and MC at the ND in the on-axis position. Fig.~\ref{fig:cdrcont} shows the result obtained from a traditional model-dependent measurement, where the on-axis only ND data-MC comparison fails to correct deficiencies in the neutrino interaction model, resulting in a biased oscillation measurement~\cite{DUNENDCDR}. The case presented in Fig.~\ref{fig:prismcontour} shows the PRISM oscillation measurement, demonstrating how an accurate, unbiased measurement is achieved by reducing dependence on the neutrino interaction model~\cite{Hasnip2023y}.

\vspace{-0.75em}
\section{\label{sec:conc}Conclusions}
\vspace{-0.75em}

Controlling the impact of systematic uncertainties is of critical importance if DUNE is to make world-leading measurements of neutrino oscillations. The PRISM system is an invaluable tool to reduce neutrino interaction model dependence and will enable DUNE to make robust measurements of neutrino oscillations. Further details on the PRISM methodology can be found in Ref.~\cite{Hasnip2023y}.

\vspace{-3em}
\printbibliography[heading=bibintoc,title={\bibtitle}]

\end{document}